\documentstyle[epsf]{article}
\def\amin{\ifmmode ^{\prime}\else$^{\prime}$\fi}
\def\asec{\ifmmode ^{\prime\prime}\else$^{\prime\prime}$\fi}
\newcommand{\be}{\begin{equation}}
\newcommand{\ee}{\end{equation}}
\newcommand{\bd}{\begin{displaymath}}
\newcommand{\ed}{\end{displaymath}}
\newcommand{\bea}{\begin{eqnarray}}
\newcommand{\eea}{\end{eqnarray}}

\newcommand{\gapprox}{\;\rlap{\lower 2.5pt
             \hbox{$\sim$}}\raise 1.5pt\hbox{$>$}\;}
\newcommand{\lapprox}{\;\rlap{\lower 2.5pt
             \hbox{$\sim$}}\raise 1.5pt\hbox{$<$}\;}

\begin{document}

\centerline{\bf Fast maximum entropy Doppler mapping}
\medskip
\centerline{H.C. Spruit}
\centerline{Max Planck Institut f\"ur Astrophysik, Box 1523, D-85740 Garching}
\centerline{henk@mpa-garching.mpg.de}
\medskip\medskip

\noindent
{\bf Abstract} A numerical code is described for constructing Doppler maps from the 
orbital variation of line profiles of (mass transfering) binaries. It uses an 
algorithm related to Richardson-Lucy iteration, and is much faster than the 
standard algorithm used for ME problems. The 
method has been tested on data of cataclysmic variables (including WZ Sge and SS 
Cyg), producing maps of up to 300X300 points. It includes an IDL-based set of 
routines for manipulating and plotting the input and output data, and can be 
downloaded from http://www.mpa-garching.mpg.de/$\sim$henk.
\medskip

\leftline{\bf Introduction}
The orbital variation of spectral line profiles in a binary contains information on 
the distribution of emitting or absorbing regions in the system. The process of 
Doppler mapping (Horne and Marsh, 1988) converts the orbital-phase dependent line 
profile into a map in velocity space. This map is easier to interpret than the 
original line profile. If the relation between velocity and position is known, for 
example if the binary parameters are known and/or the velocities in the accretion 
disk are assumed to be Keplerian, this can then be used to derive information on the 
spatial distribution of emitting material. 

Doppler mapping has become a standard tool for work on Cataclysmic Variables and 
X-ray binaries (e.g. Horne 1991, McClintock et al. 1995, Marsh and Duck 1996, 
Harlaftis et al. 1996,  1997, Schwope et al. 1997, Steeghs et al. 1997). Most 
commonly used is some form of Fourier-filtered back-projection, which is easy to 
implement and fast. It tends to produce noticeable artefacts, however, in the form of 
`radial spokes'. Much cleaner results are obtained with a Maximum Entropy procedure 
(Marsh and Horne 1988). Standard implementations of this method, however, are so slow 
that only rather small maps can be conveniently processed (interactively, say) on 
current workstations. 

The algorithm implemented here solves the same maximum entropy problem, but with a 
very much faster algorithm. Technically, it derives from the Richardson-Lucy 
maximum-likelihood algorithm, but solves a different problem. Its efficiency has been 
demonstrated before in the case of eclipse mapping (Spruit 1994). The code described here has been used by Spruit and Rutten (1998) for Doppler mapping of WZ Sge.
\medskip

\leftline{\bf Method}
The formulation of the Doppler mapping (or Doppler tomography) problem has been 
described in detail in Marsh and Horne (1988). In brief, it is assumed that the 
system consists of optically thin emission regions, and that the emission is 
stationary in a frame corotating with the orbit. The line-of-sight velocity of each 
element of emitting material then varies sinusoidally with the orbital phase. The 
line profile at each orbital phase is then a superposition a large number of these 
sinusoidally varying line profiles. In most systems, the line profile in the
rest frame of the gas is fairly narrow compared with the velocity of the gas, and is 
usually neglected campared with the Doppler shifts due to motion. 

The code described here assumes, finally, that all the emission takes place in a
single plane, the orbital plane of the system. This is appropriate if the accretion 
disk is the main object of study. In principle, however, an arbitrary geometry of the 
emitting surfaces can be included as long as it can be specified in advance. An 
example is the Roche lobe mapping of Rutten and Dhillon (1994). 

Take cartesian coordinates $v=(v_x,v_y)$ in the space describing the velocity of 
material in the orbital plane $(x,y)$. Discretize this velocity space in bins of size 
$(\Delta v_x, \Delta v_y)$ and number these points (in some fixed but arbitrary 
order) $j=1...m$. Let each of these bins have an emissivity (per unit of 
velocity$^2$) $\psi_j$, the {\it Doppler map}. This distribution of emission in 
velocity space produces a phase-dependent line profile $\varphi(v_\parallel,\phi)$, 
where $v_\parallel$ is the line of sight velocity and $\phi$ the orbital phase. Let 
$\varphi$ be discretized in $i=1...n$ points (again in arbitrary but fixed order). 
Then the transformation from the emission distribution in velocity space to the line 
profile is given by a linear matrix operator $P$:
\begin{equation} \varphi_i=\sum_j P_{ij} \psi_j. \end{equation}
The form of $P$ has been described in the previous literature cited. Let the {\it 
observed} spectrum be $\tilde\varphi_i$. The object of the reconstruction process
is now to determine a Doppler map $\psi$ such that $\varphi$ matches 
$\tilde\varphi$ within the observational uncertainties while making as few 
assumptions on the nature of the map as possible. In the `maximum entropy' 
formulation, this becomes an optimization problem in which a single scalar quantity 
$Q$ (`quality function' or `energy') is maximized:
\be
Q=H(\varphi,\tilde\varphi)+\alpha S(\psi),
\ee
where $H$, the likelihood, is a number which maximizes when the observations are 
matched exactly ($\varphi=\tilde\varphi$), and $S$, the entropy, maximizes for 
`smooth' maps. 
For the likelihood, two forms are in current use, the logarithmic one advocated by 
Lucy for positive definite images:
\be H= \sum_i\tilde\varphi_i\ln\varphi_i\ee
and a $\chi^2$-based likelihood:
\be H= -n\sum_{i=1}^n(\varphi_i-\tilde\varphi_i)^2. \ee
For the entropy, we use the form with `floating defaults' introduced by Horne (1985):
\be S=-\sum_j\left(\psi_j\ln{\psi_j\over\chi_j}+\chi_j\right).\ee
where $\chi_j$ is a {\it default image} created from the map $\psi$ by a smearing
operation. For this smearing we use a simple Gaussian smearing in velocity space with 
an adjustable width. The purpose of this default is to bias the resulting map to a 
`smooth' map without necessarily biasing it toward a completely flat one. This has 
become standard practice in eclipse mapping and Doppler mapping applications.

The value of $\alpha$ regulates how close a fit to the input data is obtained. An 
objective measure would be to choose $\alpha$ such that the reconstructed spectrum
$\tilde\varphi$ fits the input data $\varphi$ just within the error bars but not 
closer. In practice, there are always features in the input data that can not be 
fitted at all because the assumptions made (in particular the optically thin nature 
of the emission) are violated by the data. A compromise then has to be made, based in part on personal judgement.

The optimization is done iteratively by adjusting the values of $\psi$. The iteration 
is a slightly modified version of that of Lucy (1994), and is described in detail in 
Spruit (1994). It converges to a single unique maximum in 20-40 iterations, each of 
which uses 3 forward and 1 backward projections. 
\medskip

\leftline{\it Computing resources}
The computations are done in double precision. The projection matrix $P_{ij}$ 
accounts for almost all of the memory used by the code. For efficient access, it has to be stored along with its transpose. It can, however, be stored in single precision 
without loss of accuracy. Since most of the elements of $P_{ij}$ vanish, index arrays 
are used to map the indices of the nonvanishing elemnts into linear arrays. This adds 
an equal amount of storage in the form of long integers, so that the total storage is 
8 bytes per nonvanishing element of the matrix. 

Let $n_\lambda$ be the number of wavelength points of the spectra and $n_\phi$ the 
number of orbital phases (number of spectra). Let $n_v$ be the number of points in 
each of the two velocity coordinates in the Doppler map. If the resolution in the 
Doppler map is matched to the 
wavelength resolution of the spectra, $n_v\approx n_\lambda$. In this case both the 
computing time and the amount of main memory used scale as $n_v^3$, more or less 
independent of $n_\phi$. For a map of 80X80 points, produced from 160 spectra of 80 
wavelength points, the cpu time is of the order 30s, and 45 MB of memory are needed. 
The largest map on which the code was tested (from high resolution data of SS Cyg in 
quiescence) has $n_\lambda=n_v=300$, $n_\phi=300$. On the order of 15 minutes and 1.5 
GB were used.
\medskip

\leftline{\it Examples}
Fig. 1 shows the postscript output produced from a dataset consisting of 
17 spectra of 72 wavelength points. The lack of resolution in the azimuthal direction 
due to the rather low number of phase points is evident. Figure 2 shows the same 
Doppler map with mass transfering stream and secondary star overplotted. Figure 3 
shows the results from high resolution spectra of SS Cyg in quiescence. Spectral 
resolution was 0.24{\AA} (two pixels per resolution element), phase resolution 0.01. 

\medskip
\leftline{\it downloading the software}
The package can be accessed by web browser at\break{\tt 
http://www.mpa-garching.mpg.de/$\sim$henk}. It has been extensively tested on IBM AIX, 
HP/UX and Sun Solaris machines, and I consider it stable. When publishing results in 
which use has been made of this package, please acknowledge this by a reference to 
the present publication. Bug reports are welcome. The package includes a documentation 
text, sample input spectra, sample output data and and ps files from 
the graphics routines.

The actual Doppler mapping is done by a Fortran 77 program that can be used 
stand-alone. Recommended  use however is with the set of IDL routines provided. 
Apart from plotting results and input data, they are designed to simplify running of 
the Fortran code. 
\medskip

\leftline{\bf Acknowledgements}
I thank Drs. Keith Horne and Leon Lucy for discussions on various aspects of 
astromical image reconstruction problems.

\begin{figure}
\mbox{}\hfil\mbox{\epsfxsize14cm\epsfbox{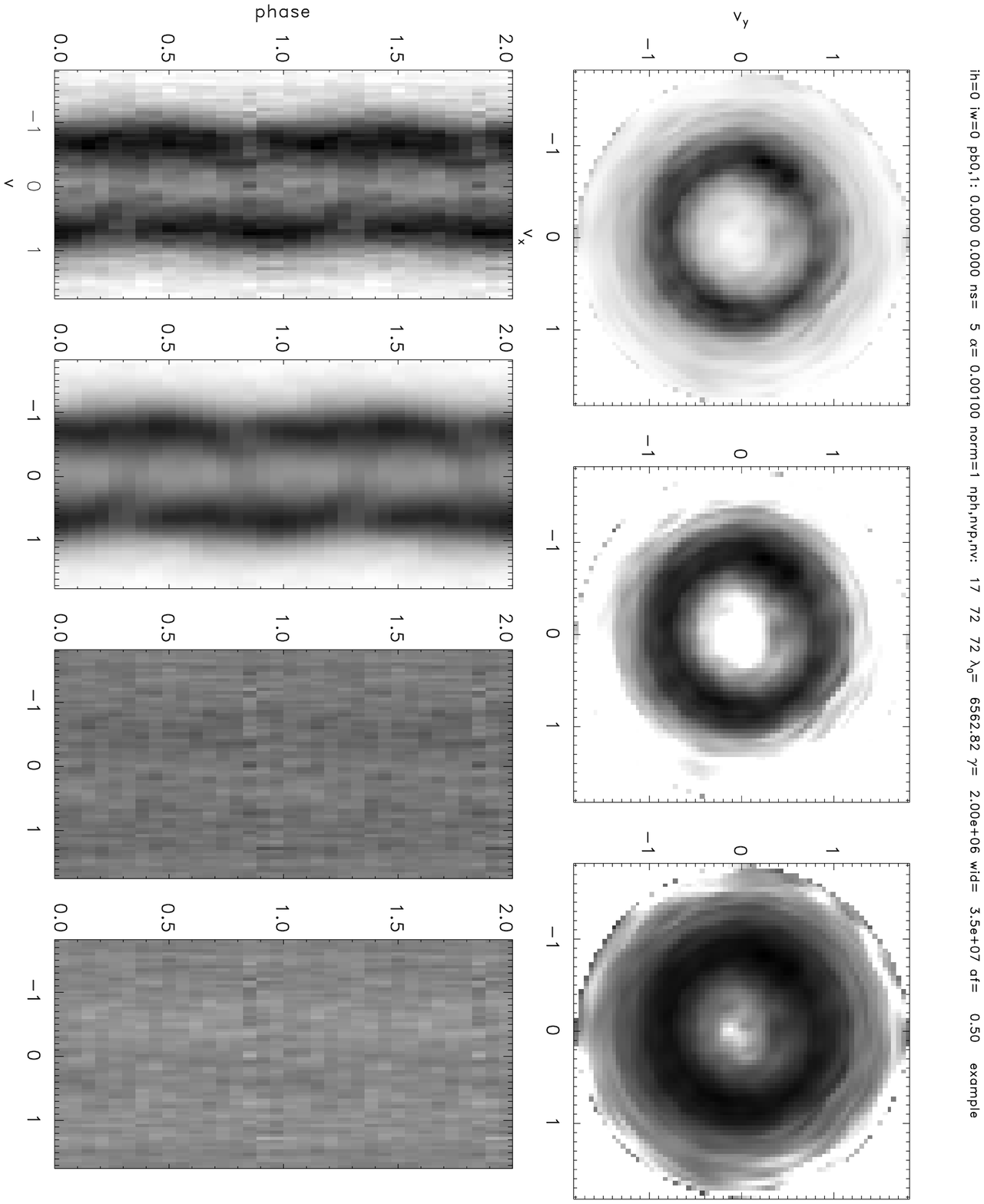}}\hfil\mbox{}
\caption{
Doppler map from observations of a CV in quiescence. 17 orbital phase points, 72 
wavelength points. Upper row: the Doppler map at 3 different scalings of the image 
intensity. Lower row, left to right: phase-folded input spectrum (O), spectrum 
reconstructed from the Doppler map (C), and the difference maps O-C and C-O 
(postscript output from IDL routine {\tt dopmap.pro}).}
\end{figure}

\begin{figure}
\mbox{}\hfil\mbox{\epsfxsize14cm\epsfbox{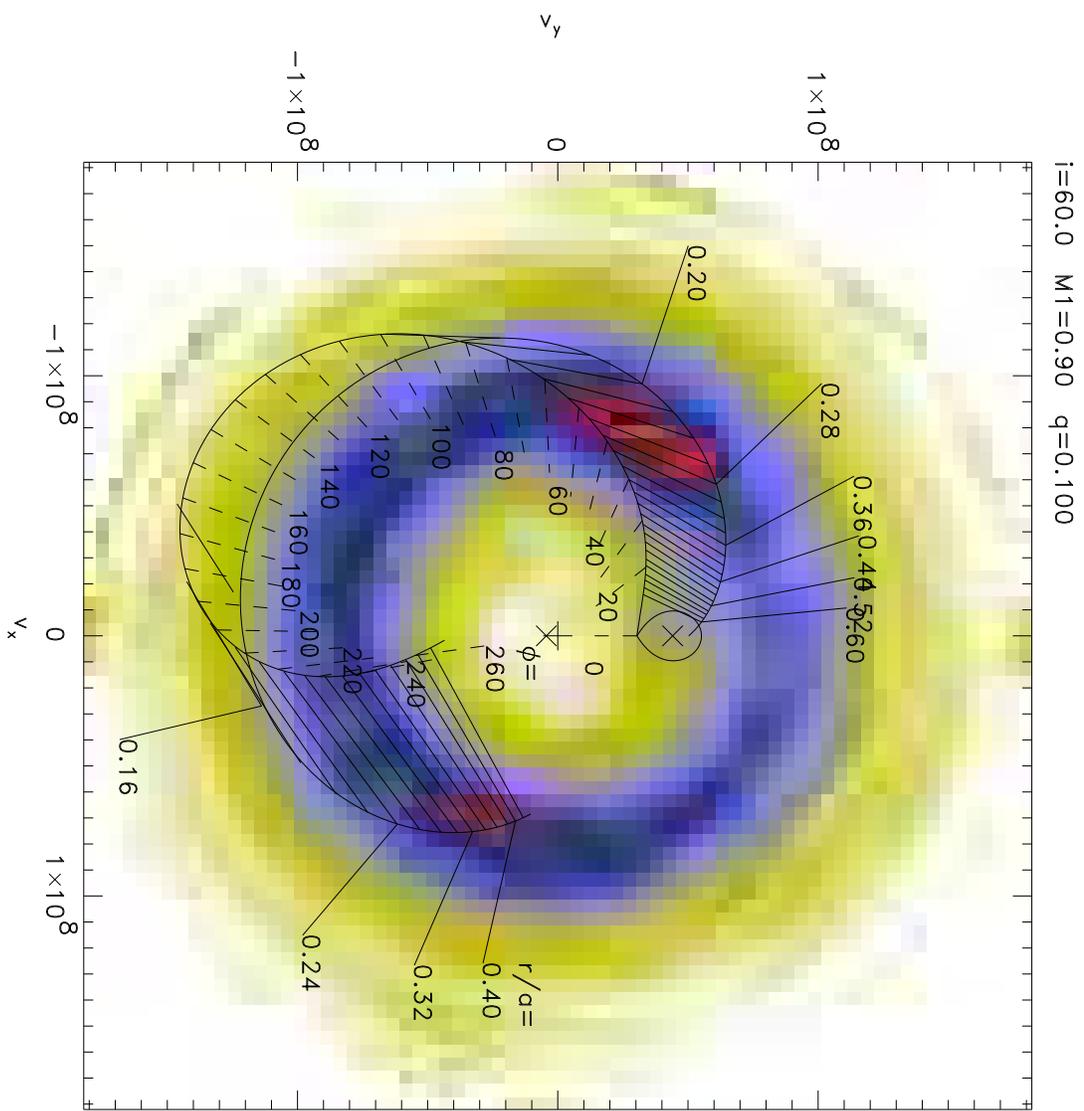}}\hfil\mbox{}
\caption{Same map as in figure 1, overplotted with the `Doppler ghost' of the 
secondary, the system center of mass (+), and the velocities of primary and secondary (X). 
The two curves show the theoretical path of the stream from L1, and the Kepler velocity 
$(GM_1/r)^{1/2}$ at the location 
of the stream. Curves are marked with the distance from the primary (in units of the 
binary separation), and the azimuth of the stream as seen from M1 (output from IDL 
routine {\tt stream.pro}).}
\end{figure}

\begin{figure}
\mbox{}\hfil\mbox{\epsfxsize14cm\epsfbox{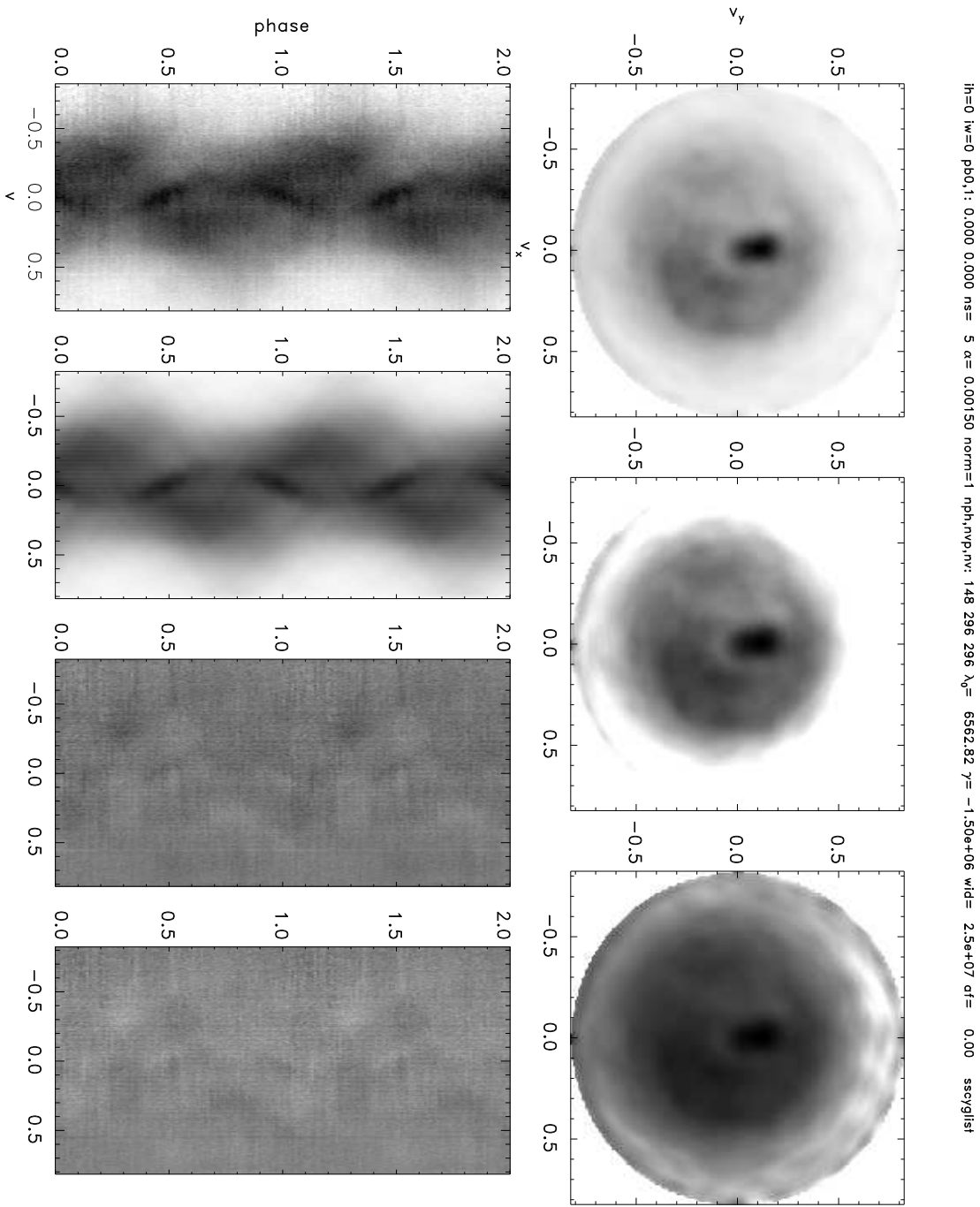}}\hfil\mbox{}
\caption{As figure 1, but for high resolution data of SS Cyg in quiescence. Phase 
resolution 0.01, 296 wavelength points.}
\end{figure}
\medskip

\end{document}